\documentclass{article}% Math and Physical Sciences Numbered Reference Style
\usepackage{graphicx}%
\usepackage{multirow}%
\usepackage{amsmath,amssymb,amsfonts}%
\usepackage{amsthm}%
\usepackage{mathrsfs}%
\usepackage[title]{appendix}%
\usepackage{xcolor}%
\usepackage{textcomp}%
\usepackage{manyfoot}%
\usepackage{booktabs}%
\usepackage{algorithm}%
\usepackage{algorithmicx}%
\usepackage{algpseudocode}%
\usepackage{listings}%
\usepackage{soul}
\usepackage{authblk}
\usepackage[a4paper, total={6in, 8in}]{geometry}

\DeclareMathOperator*{\argmin}{\arg\min}

\usepackage{tabulary}
\newcolumntype{K}[1]{>{\centering\arraybackslash}p{#1}}

\title{A Digital Twin for Microwave Liver Treatment Re-planning}

\author[1]{Ilias Nahmed\footnote{corresponding author : ilias.nahmed@inria.fr}}

\author[1]{Francesco Dettori}

\author[2]{Juan Verde}

\author[1]{Michel Duprez}

\author[1]{Pablo Alvarez}

\author[1]{Stéphane Cotin}

\affil[1]{Université de Strasbourg, CNRS, Inria, ICube, F-67000 Strasbourg, France}

\affil[2]{IHU Strasbourg, Strasbourg, France}

\date{}

\begin{document}

\maketitle

\abstract{
\textbf{Purpose:} MicroWave Ablation (MWA) modeling and simulation bear great potential for loco-regional treatment of liver tumors. However, accurately positioning the antenna according to a planned orientation/location is technically challenging. In cases of misplacement, maintaining the original plan may cause incomplete ablation, while repositioning the antenna may induce tumor seeding. In this work, we propose (i) a digital twin of MWA that simulates ablation outcomes, and (ii) an optimizer that suggests corrections to MWA parameters without antenna reinsertion, while ensuring complete tumor ablations.

\textbf{Methods:} A finite element scheme was used to solve the coupled microwave propagation and heat transfer equations governing MWA, with personalized dielectric and thermal properties determined from preoperative CT and MRI images. We then proposed an optimization algorithm able to adjust power input, ablation duration, and antenna position to correct for antenna misplacement.

\textbf{Results:} The simulator and optimizer were evaluated against \textit{in vivo} swine experimental data. Three ablations were performed in liver regions with varying vascularization. The simulations accurately predicted the ablation zones despite the presence of large vessels near the antenna, achieving Dice scores of $0.82$, $0.81$, and $0.79$. In the case of replanning scenarios, our optimizer predicted new parameter sets that led to Dice scores of 0.83, 0.83, 0.80, a corresponding improvement of 20.3\%, 40.7\% and 48.1\% in average over the initial ablation result.

\textbf{Conclusion:} This paper is the first to address intra-operative replanning of thermal ablation therapy. It demonstrates that optimal ablation results can be achieved without requiring antenna reinsertion by optimizing specific ablation parameters.
}

% \keywords{Microwave ablation, Digital twin, Simulation, Planning optimization}

\section{Introduction}\label{sec1}
Liver cancer is the third leading cause of cancer-related death in the world \cite{tu_current_2024}. Although surgical resection and liver transplantation are established as effective treatment options, they remain unsuitable for more than 80\% of the patients \cite{poulou_percutaneous_2015}. For these reasons, along with reduced costs and improved repeatability, minimally invasive locoregional treatments based on thermal ablation have gained popularity in the last decades \cite{simon_microwave_2005}. These techniques also enforce parenchyma preservation, particularly when treating small tumors.

Microwave ablation (MWA) constitutes, to date, one of the most effective thermal ablation modalities among the clinically available options \cite{sugimoto_microwave_2025}. MWA operates through a process known as dielectric hysteresis, where heat is generated from the friction and collision of water molecules (inside biological tissues) as they spin to align with the rapidly changing electromagnetic field propagating from the applicator antenna \cite{simon_microwave_2005}. In comparison with other thermal ablation techniques, MWA requires shorter times to produce higher temperature profiles, generally leading to larger ablation zones that can be used to treat larger tumors safely. Moreover, MWA is less susceptible to the heat-sink effect, where ablations in proximity to blood vessels are less efficient due to temperature losses through advection \cite{sugimoto_microwave_2025, simon_microwave_2005}.

Treatment through MWA can only be successful under appropriate preoperative planning. However, MWA planning in current clinical practice relies on geometrical models based on power \textit{versus} ablation-duration charts provided by antenna manufacturers. Since these geometrical models completely disregard patient-specificity, they generally overestimate the ablation zones, leading potentially to incomplete ablations that could cause tumor recurrence \cite{Winokur2014}. With the emergence of computational models, it has become possible to simulate numerically the physical process of MWA in biological tissue. However, MWA involves various physical phenomena, including electromagnetics, heat transfer, and computation of tissue damage. Several methods, like Finite Elements (FE), Finite Differences, and Lattice Boltzmann, constitute suitable numerical frameworks for solving such multiphysics problems \cite{audigier:tel-01256010,singh_thermal_2020,boskovic_finite_2023}. Yet, to achieve optimal ablation outcome prediction, it is essential to account for patient-specific tissue properties \cite{servin_simulation_2024, heshmat_using_2024}.

Once a plan has been established -- via numerical simulation or otherwise -- the antenna is inserted into the liver under CT guidance. However, because of breathing motion, changes in patient position, and insertion-induced deformation, it is practically impossible to position the antenna as planned \cite{seitel_reinsertion,heshmat_using_2024}. This is particularly true when multiple antennas are required to treat larger tumors.
Thus, conforming to the preoperative plan often necessitates repositioning the antenna, possibly through reinsertion, which is both time-consuming and dangerous. Indeed, puncturing the tumor once, let alone several times, can lead to tumor seeding \cite{patel_no-touch_2013}. Fortunately, likely, an appropriate ablation zone can still be produced by fine-tuning ablation power and duration parameters, while avoiding both antenna reinsertion and excessive damage to healthy tissue. One can also consider slight displacements along the insertion axis, which is reasonably safer than antenna reinsertion and repositioning. As these adjustments are typically determined empirically, a re-planning tool able to quantitatively compute them could offer substantial clinical advantages.

{In this paper, we present a patient-specific simulation of MWA that can be used to simulate and optimize the therapy. Sec. \ref{sec:simulation} presents the computational model, while Sec. \ref{sec:antenna} and \ref{sec:parametrization} show its parametrization. The presented model is validated in Sec. \ref{sec:ablation-results}.  In Sec. \ref{sec:replanning} we address the (re)planning solution, and present our results in Sec. \ref{sec:replanning-results}.}

\section{Microwave ablation simulation}

In this section, we describe how heat, generated by the MWA antenna, diffuses across soft tissues. Then, we explain how the model's parameters can be tuned to characterize both the antenna and the organ, making our solution patient-specific. 

\subsection{{Computational model}} \label{sec:simulation}

The Pennes Bioheat equation is the most widely used model for simulating heat transfer within tissues following thermal ablation \cite{singh_thermal_2020}. The governing equation of this model is given by:
\begin{equation}
   \rho C \frac{\partial T}{\partial t}\ =\ \nabla\cdot(\kappa \nabla T)\ - \ \rho_b C_b \omega_b (T-T_b) + Q_m + Q_{ext},
   \label{eq:pennes}
\end{equation}

where $\rho$, $C$, $\kappa$ and $T$ are the tissue density (kg/m$^3$), specific heat capacity (J/kg/K), thermal conductivity (W/m/K) and temperature (K) of the tissue, respectively; $\rho_b$, $C_b$ and $T_b$ correspond to the same physical quantities for the blood entering the tissue, $\omega_b$ is the blood perfusion rate (1/s) that allows modeling the heat sink effect \cite{singh_thermal_2020}, $Q_m$ is the metabolic heat source and $Q_{ext}$ is the external heat source.  

A critical factor in MWA modeling is the temperature dependency of tissue parameters. Indeed, the thermal effects of MWA induce essential changes in tissue (\textit{e.g.} water evaporation, coagulation, cellular denaturation), which alter both the microwave energy deposition and heat transfer phenomena. Following previous works, this work accounted for tissue coagulation in blood perfusion $\omega_b$ \cite{ge_multi-slot_2018, Zhu2013}, a thermal conductivity $\kappa$ linearly increasing with temperature \cite{deshazer_physical_2016}, as well as a water-content dependency for the heat capacity $C$ \cite{cavagnaro_numerical_2015}.

In \cite{ge_multi-slot_2018, Zhu2013}, perfusion shutdown is modeled by a non-differentiable temperature threshold, which leads to difficulties in numerical solvers. To preserve the same physical behavior while ensuring differentiability, we replace the sharp switch by a smooth approximation
\begin{equation}
\begin{aligned}
\omega_b^{\text{sharp}}(T) &=
\begin{cases}
\omega_0, & T<60^\circ\mathrm{C},\\
0, & T\geq60^\circ\mathrm{C},
\end{cases}
\qquad
\omega_b^{\text{smooth}}(T)
= \omega_0\left(\tfrac{1}{2}\tanh(60-T)+\tfrac{1}{2}\right),
\end{aligned}
\end{equation}
where $\omega_0$ denotes the blood perfusion at $37^\circ\mathrm{C}$

The metabolic heat source $Q_m$ is usually neglected in thermal ablation modeling, since it is tiny in comparison to the external heat source $Q_{ext}$. The latter models heating as microwaves propagate through tissue, and is defined in relation to the specific absorption rate (SAR) by
\begin{equation}
     Q_{ext} = \rho \cdot \mathrm{SAR} = \frac{1}{2} \sigma\langle\vec{E}, \vec{E}^*\rangle,
\end{equation}
with $\vec{E}$ the solution to the microwave propagation problem described by the frequency-domain Maxwell equation:   
\begin{equation} \label{eq:maxwell}
\nabla \times \mu_r^{-1}  (\nabla \times \vec{E})  - k_0^2(\varepsilon_r - j\frac{\sigma}{\omega \varepsilon_0}) \vec{E} = 0,
\end{equation}
where $\sigma$ is the electric conductivity (S/m), $\mu_r$ is the relative permeability, $\varepsilon_r$ is the relative permittivity, $\omega = 2\pi f$ is the angular frequency (rad/s) for $f = 2450~\text{MHz}$, $\varepsilon_0 = 8.854\times 10^{-12}~\text{F/m}$ is the permittivity in vacuum and $k_0 = 51.35~\text{m}^{-1}$ is the propagation constant in vacuum, respectively.

Similar to the thermal properties in Eq.~\eqref{eq:pennes}, dielectric properties also vary with temperature \cite{singh_thermal_2020}. Previous works have either neglected this dependency \cite{servin_simulation_2024}, or solved the coupled system of equations using time-discretization schemes of varying degrees of complexity and stability \cite{boskovic_finite_2023, cavagnaro_numerical_2015}. Arguably, the most critical consequence of this temperature dependency is the decay of the external heat source $Q_{ext}$ with increasing temperature. Therefore, in the interest of saving computational time while maintaining the temperature dependency, the effective external heat source is approximated by: 
\begin{equation}
    Q_{ext}(T) = \left(1 - \frac{1}{1+\exp(6.583-0.0598\cdot T)} \right) \frac{\sigma}{2} \langle\vec{E}, \vec{E}^*\rangle,
    \label{eq:heat_source}
\end{equation}
where the exponentially decaying factor is adopted from the temperature-dependent electric conductivity model in \cite{boskovic_finite_2023}. 

In Eq.~\eqref{eq:maxwell}, the power input from the generator is modeled by a port excitation boundary condition at the entry point of the antenna's coaxial waveguide, the conducting materials of the antenna are idealized and modeled by Perfect Electrical Conductor boundary conditions at their surfaces, and a first-order scattering boundary condition is applied at the remaining outer domain boundaries to avoid reflection of electromagnetic waves. As for Eq.~\eqref{eq:pennes}, all boundaries were considered thermally insulated, except for the antenna/tissue interface, where a convection boundary condition was applied to model the antenna's cooling technology \cite{cavagnaro_numerical_2015}. The validity of the first-order scattering boundary condition was verified by a domain-size convergence study of the electromagnetic problem, showing that boundary reflections become negligible near the ablation zone for sufficiently large domains. In particular, beyond a transverse domain width of 20~mm, the relative electric field $L^2$ error remained below 0.5\%, indicating that residual boundary artifacts do not affect the predicted ablation shape.

A finite element numerical scheme was implemented using the open source library FEniCSx\footnote{https://fenicsproject.org/} to solve the coupled microwave propagation and heat transfer equations governing MWA. Axial symmetry was assumed for the microwave propagation problem, since low variability of the electric field solution to variations in dielectric properties has been reported in previous work \cite{deshazer_physical_2016}. Thanks to the temperature-dependent external heat source $Q_{ext}$ introduced in Eq.~\eqref{eq:heat_source}, the microwave propagation and heat transfer problems are only weakly coupled, and a single resolution of the linear system in Eq.~\eqref{eq:maxwell} is required. 
The validity of this approximation was quantitatively verified by comparison with a strongly coupled formulation, in which the Maxwell equation is re-solved with updated dielectric properties at each thermal iteration. The resulting differences in temperature and ablation zones were found to be negligible across a wide range of blood perfusion values, with a maximal temperature RMSE below $0.83^\circ\text{C}$ and a maximal Arrhenius factor RMSE below $1.6\times10^{-2}$ units. In addition, comparisons between heterogeneous and homogeneous electromagnetic models, as well as between 3D and interpolated 2D electromagnetic solutions, showed that tissue heterogeneity and dimensional reduction introduce only minor errors in the temperature and damage, achieving RMSE values as small as $1,07^\circ\text{C}$ for the temperature, and $ 6.6 \times 10^{-2}$ units for the damage. These results justify the proposed weakly coupled, axisymmetric electromagnetic modeling, which captures the dominant heating mechanisms while significantly reducing computational cost.  
An implicit time-discretization scheme was implemented to solve the nonlinear system in Eq.~\eqref{eq:pennes}, requiring a Newton resolution per time increment ($dt = 10\mathrm{s}$).

\subsection{Antenna characterization}
\label{sec:antenna}
The geometry of the antenna strongly influences the pattern of the generated electromagnetic field, and thus the specific absorption rate, which dictates the resulting extent and shape of the ablation zone. The precise design of MWA antennas is, unfortunately, generally unknown, besides their external geometry. It is therefore essential to estimate the antenna characteristics as accurately as possible.

Since MWA antennas are designed to optimize the $S_{11}$ reflection coefficient (which measures the efficiency of energy delivery), the unknown antenna geometry parameters were estimated through a constrained optimization problem. The objective was to minimize the 
the reflection coefficient $S_{11}$, by varying the antenna geometry while accounting for estimates of some of its characteristics: quarter wavelength choke, 50~Ohm impedance inner/outer conductor radii, total radius. The specifics of this optimization process are out of the scope of this paper, but interested readers are referred to \cite{Etoz2017} for further details. 

\subsection{Patient-specific parametrization}
\label{sec:parametrization}

This work relies on a patient-specific, multi-compartment anatomical model built from preoperative medical images. 
Experimental data were acquired to personalize and validate our MWA simulations. All preclinical data were obtained from \textit{in vivo} experiments on swine models following a protocol closely aligned with the standard of care. A pre-treatment contrast-enhanced CT (CE-CT) scan was performed to identify the target ablation location and extract the swine anatomy for material property assignment.

For the microwave propagation problem, all the patient-agnostic parameters related to the antenna materials and liver tissue were obtained from publicly available databases \cite{Baumgartner2025}. 
For the heat transfer problem, the thermal properties of each tissue type were derived from the literature \cite{vaidya_tuning_2022, deshazer_physical_2016, Baumgartner2025} and assigned to the corresponding anatomical compartments, as reported in Table~\ref{tab:mat_props}.
Particular attention was given to the blood perfusion parameter $\omega_0$, since it greatly influences the solution of the Pennes Bioheat equation. Although a constant value is typically used for the vessels, a wide range of values for the healthy liver tissue can be found in the literature \cite{deshazer_physical_2016, audigier:tel-01256010, Baumgartner2025, heshmat_using_2024}, from $0.015 \text{s}^{-1}$ to $0.071 \text{s}^{-1}$. This variability can lead to ablation volume differences as large as $18$~cm$^3$.
To address this issue, a calibration of the parameter $\omega_0$ was performed on a baseline ablation far from the large vessels using the same objective function as the one later employed in the optimization problem described in Sec. \ref{sec_opt_replan}.
A post-treatment multiphase CE-CT scan was then collected to assess this baseline ablation. The antenna was left in place during imaging, ensuring its visibility in the post-ablation scan.
This patient-specific estimated value of $\omega_0$ was then used in all following simulations. 

\begin{table}[h]
\centering
\caption{Multi-compartment material properties for thermal transfer problem \label{tab:mat_props}}
\begin{tabular}{@{}ccccc@{}}
\toprule
Tissue type     & $\kappa_0$ [W/m/K] & $C_0$ [J/kg/K] & $\rho$ [kg/m$^{3}$] & $\omega_0$ [1/s] \\ 
\midrule
Liver Tissue   & $0.52$  & $3540$     & $1080$ & $0.079$ (fitted value) \\ 
Blood vessels   & $0.54$ &  $3770$    & $1060$  & $0.2$  \\ 
\bottomrule
\end{tabular}
\end{table}

\subsection{Ablation simulation results}
\label{sec:ablation-results}

We validated our results using \textit{in vivo} swine experiments with three ablations scenarios at the IHU of Strasbourg, with the approval of the local ethics committee. 
Although limited to three ablations, these \textit{in vivo} experiments are intended as a qualitative validation of the proposed modeling framework under realistic physiological conditions, and not as a statistical or population-level study.
The three ablations ($100~\text{W}$ during $5$ minutes) were performed in regions with different degrees of vascularization: 1) away from any big vessels to calibrate the $\omega_0$ parameter and to constitute a baseline (labeled \textit{B}); 2) near the hepatic vein \textit{(HV)} and 3) near both hepatic and portal veins \textit{(HPV)}. This choice was made to estimate robustness to the heat sink effect. The antenna used in this study was a 15-gauge \textit{Medtronic Covidien Emprint™} antenna with water-cooling technology, delivering up to 150W at a frequency of 2.45GHz.

The solution of the bioheat equation \eqref{eq:pennes} is a time-dependent temperature field, which can only be measured experimentally using MR thermometry. This imaging modality is rarely available and prone to measurement errors. As an alternative, we evaluate our results against the actual ablation region. To achieve this, we simulate the ablation using the Arrhenius thermal injury model: 
\begin{equation}
\label{eq:arrhenius}
    \Omega (t) = \int_0^t A \exp \Big(\frac{-\Delta E}{RT(\tau)}\Big)d\tau ,
\end{equation}
where $A = 7.39 \times 10^{39} \text{s}^{-1}$ is a frequency factor, $\Delta E = 2.577\times 10^5 \text{J mol}^{-1}$ is the activation energy, $R = 8.314 \text{J mol}^{-1}\text{K}^{-1}$ is the universal gas constant and $T(\tau)$ is the temperature at time $\tau$ \cite{servin_simulation_2024}. The cell death probability $\theta_d$ is then computed as $ \theta_d = 1 - e^{-\Omega(t)} $ where the threshold $\theta_d > 0.98$ is used for indicating cell necrosis \cite{servin_simulation_2024}.

\begin{figure}
    \centering
    \includegraphics[width=1\linewidth]{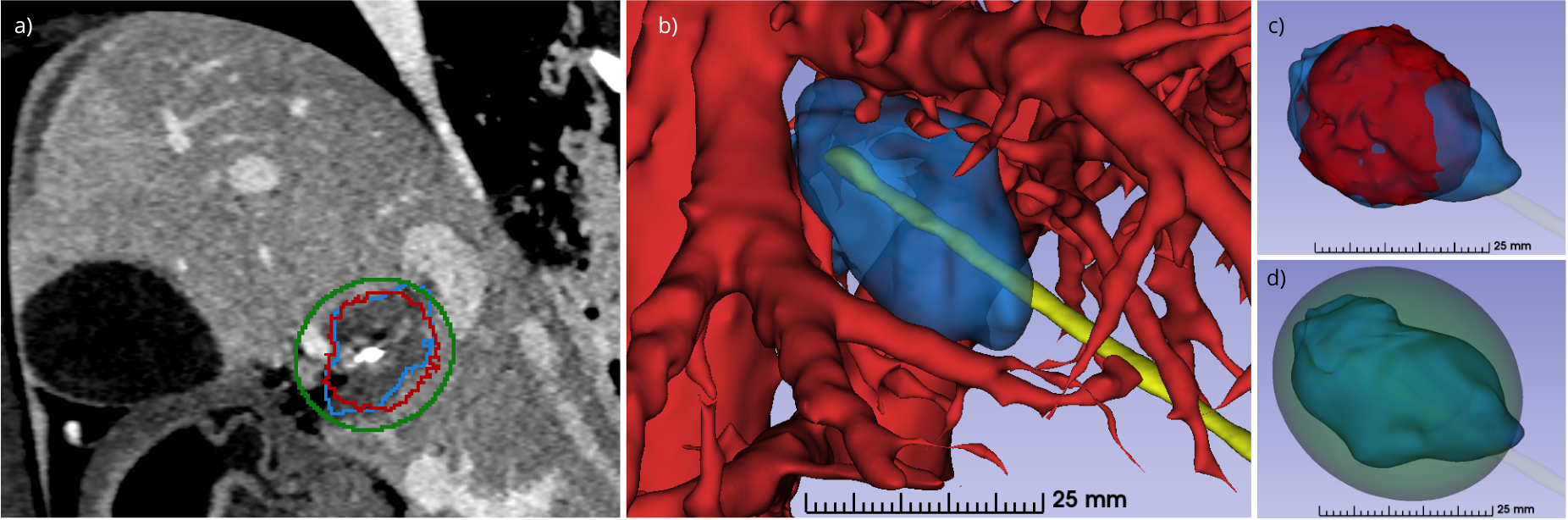}
    \caption{Visual and quantitative comparison of ablation segmentation models for the HPV case. (a) CT slice with outlines: ground truth (blue), our prediction (red), and manufacturer's prediction (green). (b) 3D visualization of the ground truth ablation (blue) with surrounding vasculature (red). (c) Our model's prediction (red) vs. ground truth (blue) (Dice = 0.79). (d) Manufacturer's model (green) vs. ground truth (blue) (Dice = 0.58)}
    \label{fig:comparison}
\end{figure}

All experimental ablations were replicated via simulation: same antenna location, power input (100~W), and ablation duration (5 min). The predicted ablation zones, as per Eq.~\eqref{eq:arrhenius}, were compared with the post-treatment CE-CT segmentations to measure their accuracy.
Fig. \ref{fig:comparison} presents qualitative results for the HPV ablation case, with 2D and 3D comparisons of the ground truth ablation zone, against the predicted ablation zones from our model and the manufacturer's geometrical model.
Table \ref{table3} summarizes the quantitative results of our validation:
\begin{table}[h!]
\centering
\caption{Dice similar coefficient ($\text{D}_{\text{sim.}}$, $\text{D}_{\text{man.}}$), standard Hausdorff distances ($d^{H}_{\text{sim.}}$, $d^{H}_{\text{man.}}$), and Hausdorff distances with $95^{\text{th}}$ percentile ($d^{H}_{95,\text{sim.}}$, $d^{H}_{95,\text{man.}}$) for the simulated and manufacturer's ablation zones with respect to \textit{in vivo} experimental ground truth for the ablations B (baseline far from large veins), HV (near the hepatic vein) and HPV (near the hepatic and portal vein) \label{table3} }
\begin{tabular}{lK{1.5cm}K{1.5cm}K{1.5cm}K{1.5cm}K{1.5cm}K{1.5cm}}
\toprule 
& $\text{D}_{\text{sim.}}$ & $d^{H}_{95,\text{sim.}}$ & $d^{H}_{\text{sim.}}$ & $\text{D}_{\text{man.}}$ & $d^{H}_{95, \text{man.}}$  & $d^{H}_{\text{man.}}$  \\
\midrule 
B & {0.82} & {1.7} mm & {5.9} mm & 0.62 & 4.7 mm & 8.0 mm \\
HV & {0.81} & {2.8} mm & 8.1 mm & 0.66 & 4.2 mm & {7.3} mm \\
HPV & {0.79} & {2.7} mm & {6.4} mm & 0.58 & 5.5 mm & 8.2 mm \\

\bottomrule 
\end{tabular}
\end{table}

Our simulator outperformed the manufacturer's prediction across all ablations. It achieved substantial Dice coefficient improvements of 32\% (B ablation), 22\% (HV ablation), and 36\% (HPV ablation), while the $d^{H}_{95}$ distance was reduced by 2.4mm on average. 
The simulator achieved lower $d^{H}$ values than the antenna manufacturer's model on all ablations but one (HV); however, the simulator's $d^{H}_{95}$ being lower on this ablation suggests that this difference is likely due to an outlier. In fact, $d^{H}$ is more sensitive to isolated boundary deviations than $d^{H}_{95}$.

Overall, the Dice coefficients were high (around 0.8), and the $d^{H}_{95}$ values remained very low (around 2mm), indicating clinically acceptable precision. Even with the heat sink effect present in the HV and HPV ablations, the simulator’s performance was not notably reduced: both D and $d^{H}_{95}$ values remained within the same range as the baseline ablation. These results suggest that the simulator robustly models perfusion, and the combination of high Dice scores and low $d^{H}_{95}$ values supports the validity of the simulation model.

\section{Intervention replanning through optimization}\label{sec_opt_replan}

In this section, we propose an intraoperative strategy to adjust specific ablation parameters — namely, power, ablation duration, and electrode translation — in the event of electrode misplacement.

\subsection{Ablation replanning method} \label{sec:replanning}

Placing the antenna at the intended position and orientation constitutes a significant technical challenge. Indeed, the liver may have deformed significantly during antenna insertion with respect to the preoperative plan, and this is due to the antenna itself, breathing, and/or patient repositioning. Consequently, it is not unrealistic to consider that, in many cases, the antenna position and orientation deviate from their intended positions after an initial insertion. 

\begin{figure}[htb]
    \centering
    \includegraphics[width=0.9\linewidth]{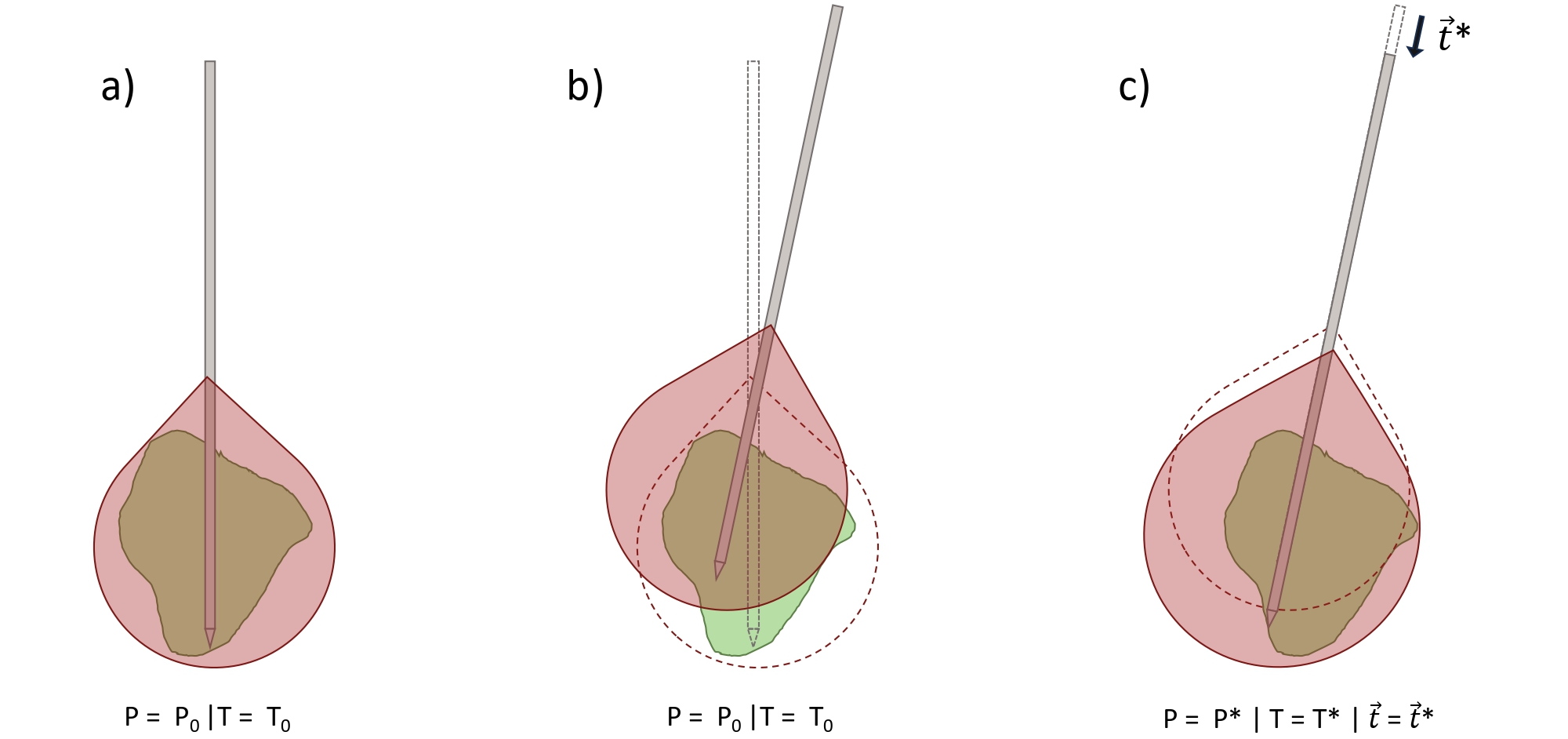}
    \caption{Illustration of the replanning workflow. (a) The ideal initial plan ($P_0$,$T_0$) achieves full tumor coverage (b) Using the same configuration when the antenna is misplaced may cause incomplete tumor coverage  (c) The proposed replanning tool suggests a translation $\vec{t}^*$ of the antenna along the same axis with updated power $P^*$ and duration $T^*$ to ensure complete ablation }
    \label{fig:replanning}
\end{figure}

Considering this antenna misplacement context, the strategy introduced here allows the computation of a new ablation plan, while avoiding the complete withdrawal of the antenna that would otherwise introduce risks of tumor seeding. Specifically, it determines an optimal translation $\vec{t}^*$ along the antenna's shaft, together with an updated power $P^*$ and ablation duration $T^*$ -- within realistic clinical bounds --, as to achieve a complete tumor ablation while minimizing damage to healthy tissue. The general workflow for the proposed MWA replanning strategy is illustrated in Fig.~\ref{fig:replanning}. 

For a new ablation plan to be admissible, the simulated ablation zone should match as closely as possible the originally planned target ablation zone. 
In this work, this was achieved by maximizing the Dice similarity coefficient, denoted by $D$, while simultaneously minimizing a modified Hausdorff distance \cite{reinke2023commonlimitationsimageprocessing}, denoted by $d^{H}_{95}$, between the simulated and target ablation zones. More precisely, our replanning strategy finds a new set of optimal parameters $P^*$, $T^*$, $\vec{t}^*$ via the following minimization problem
\begin{equation}
\label{eq:optim_replanning}
  \argmin_{P,T,\vec{t}} \frac{1}{2} \left(1 - \text{D}\left(A_{\text{sim}}\left(P,T,\vec{t}\;\right), A_{\text{target}}\right)\right) + \frac{1}{2} d^{H}_{95}\left(A_{\text{sim}}\left(P,T,\vec{t}\;\right), A_{\text{target}}\right),
\end{equation}
where $A_{\text{sim}}(P,T,\vec{t}\,)$ denotes the simulated ablation zone corresponding to a given power $P$, ablation duration $T$, antenna translation along its axis $\vec{t}$; and $A_{\text{target}}$ is the target ablation zone. 

In addition, some clinically relevant constraints were imposed : a) the power input must remain within $20~\text{W}$ and $150~\text{W}$; b) the ablation duration is limited to $15$ minutes; and c) the antenna translation along its axis is restricted to $[-5,5]~\text{cm}$. 
Because the optimization problem is highly non-linear and lacks sufficient regularity for gradient-based algorithms, genetic algorithm was employed \cite{Michalewicz1996EvolutionaryAF}. The population size was set to 15, and the maximum number of iterations to 100, with an average optimization iteration of 55 seconds, the total computation time for the optimization algorithm ranged between 50 and 95 minutes. This computational cost reflects a proof-of-concept re-planning framework and is not intended to demonstrate real-time or intra-operative feasibility, which would require surrogate or reduced-order models beyond the scope of this work.

\subsection{Ablation replanning results}
\label{sec:replanning-results}

The most natural way of validating the proposed replanning strategy would be to re-plan an ablation after a failed attempt of antenna positioning in the tumor, as illustrated in Fig.~\ref{fig:replanning}. However, reproducing such scenario in swine models (which generally do not have tumors) was not possible. As an alternative, we conducted experiments on a healthy pig, and considered the ground-truth ablated zones directly as the target ablation zones $A_{\text{target}}$ in \eqref{eq:optim_replanning}. The optimizer was then evaluated by providing it with an antenna placement intentionally shifted from the true experimental position.

Under these considerations, a perfect optimization should find a new plan (power/time/translation) that (i) corrects the initial translation, and (ii) predicts an ablation zone identical to the target ablation zone. The Dice coefficient between these two ablation zones would then be a solid way to evaluate the replanning strategy. However, the optimization problem \eqref{eq:optim_replanning} is ill-posed, since several sets of parameters $(P, T, \vec{t}\,)$ could yield the same simulated ablation zone $A_{\text{sim}}$. Although this issue has no significance in a real scenario where the objective is tailored towards the target ablation zone $A_{\text{target}}$ (any couple of parameters is admissible), it does introduce a bias in the evaluation of the proposed ablation replanning strategy. For this reason, the ablation duration $T$ was fixed at 5 min during replanning, as in the experimental ablations, and only the power input $P$ and translation $\vec{t}$ were optimized. 

The antenna positions from the three performed experimental ablations were artificially shifted by 5~mm along the antenna axis. Hence, the optimal ``new'' plan should correspond to the original power input ($P=100$~W) and the correct translation shift ($\vec{t}=-5$~mm). For each ablation, the optimization was run 50 times, each with a different random seed for the genetic algorithm in order to assess optimizer robustness. Average values and standard deviations of the Dice coefficient, Hausdorff distances, and relative errors (normalized by permissible ranges: 20:150W for power, -50:50mm for translation) in power and translation predictions are listed in Table~\ref{table4}.

\begin{table}[h!]
\centering
\caption{Dice similarity coefficient $\text{D}$, standard Hausdorff distances $d^{H}$, and Hausdorff distances with $95^{\text{th}}$ percentile $d^{H}_{95}$ for the proposed ablation replanning strategy for the ablations B (baseline far from large veins), HV (near the hepatic vein), and HPV (near the hepatic and portal vein) $\text{D}_{\text{initial}}$ corresponds to Dice coefficient without replanning. \label{table4}}
\begin{tabular}{lcccccc}
\toprule 
& $\text{D}$ & $d^{H}_{95}$ & $d^{H}$ & Power relative error & Trans. relative error  &$\text{D}_{\text{initial}}$ \\
\midrule 
B & $0.83\pm 0.002$ & $1.67$ $\pm 0.1$ mm & $5.2$ $\pm 0.5$ mm & $2.3\%$ $\pm 0.0$ & $0.5\%$ $\pm 0.0$  & 0.69 \\
HV & $0.83\pm 0.003$ & $1.63 $ $\pm 0.1$ mm & $5.7$ $\pm 0.5$ mm & $2.3\%$ $\pm 0.0$  & $2.1\%$ $\pm 0.0$  & 0.59  \\
HPV & $0.80 \pm 0.015$ & $2.17$ $\pm 0.3$ mm & $6.3$ $\pm 0.8$ mm & $4.7\%$ $\pm 0.0$  & $1.1\% $ $\pm 0.0$  & 0.54 \\
\bottomrule 
\end{tabular}
\end{table}

The optimizer recovered the target translation and power input with acceptable accuracy across all three ablations. The Dice coefficients ranged from 0.80 to 0.83, improving over the ablation predictions without replanning by 36.6\% in average.

For the translation parameter, the optimizer achieved high accuracy for ablation B, with a mean relative error of 0.5\%. Ablations HV and HPV exhibited slightly larger errors, with relative translation errors of 2.1\% and 1.1\%, respectively. Despite the increased vascular complexity in these scenarios, the variability remained negligible, suggesting stable convergence toward clinically acceptable solutions.

Power estimation showed similarly stable behavior. Relative power errors were limited to approximately 2.3\% for ablations B and HV, and increased to 4.7\% for HPV. The higher error observed in the HPV case is consistent with the presence of multiple large vessels in close proximity to the target, which increases the sensitivity of the thermal field to power variations. Nevertheless, the high Dice coefficients and low Hausdorff distances indicate that the optimizer successfully identifies power–translation combinations that compensate for these effects and accurately reproduce the ground-truth ablation shapes.

\section{Conclusion}
In this study, we introduced a patient-specific MWA simulator that accurately predicts ablation zones from \textit{in vivo} experiments with Dice scores ranging from $0.79$ to $0.82$, and Hausdorff distances of approximately $2\ \text{mm}$. In comparison with the ablation geometrical predictions provided by the antenna manufacturer, our simulations demonstrated consistently higher accuracy, especially in challenging situations where the ablations are targeted in the vicinity of large vessels, such as the hepatic or portal veins. 
Furthermore, a replanning optimizer that could be used intra-operatively — to the best of our knowledge, the first of its kind — was presented and validated using \textit{in vivo} ablations. In most clinical settings, it is almost impossible to replicate the planned antenna position exactly. The presented optimizer introduces a potential intraoperative correction for suboptimal or failed plans, which could achieve complete tumor ablation with high accuracy. This optimizer was validated using ablations on a healthy animal and achieved high accuracy and low parameter-recovery errors.

There is one limitation to the current forward simulation approach: the calibration of the $\omega_b$ parameter requires an ablation to be performed beforehand. While this was feasible in our swine experiments, it would not be the case in a clinical setting. Fortunately, several approaches exist that estimate the $\omega_0$ parameter directly from CT images. Such methods would personalize blood perfusion values, but they were not investigated in this work, as they fall outside the scope of this study. 

The simulator presented in this paper can already serve as an effective planning tool in simple clinical settings. Its future applications are up-and-coming, as it could serve as a basis for planning more complex interventions. For instance, completely avoiding tumor seeding through no-touch techniques is a promising perspective, which would require thorough planning to account for multiple impacts and necrosis coagulation. 

Regarding the replanning tool, we identified several areas for improvement. First, the current method does not account for heterogeneous tissues, in particular, the presence of a tumor. An upcoming clinical study will bring an opportunity to develop further and validate our method. Further, the computation time could be improved through the use of GPUs or surrogate deep learning models. 

\section*{Declarations}

\begin{itemize}
\item Funding: This work was funded by the MEDITWIN Bpifrance i-Demo project as part of the France 2030 program.
\item Conflict of interest: The authors have no conflict of interest to declare.
\item Code availability: The code and data supporting this study are not publicly available at this stage of the work.
\end{itemize}

\bibliographystyle{abbrv}
\bibliography{IPCAI2026}

@Article{Etoz2017,
  author    = {Etoz, Sevde and Brace, Christopher L.},
  journal   = {International Journal of RF and Microwave Computer-Aided Engineering},
  title     = {Analysis of microwave ablation antenna optimization techniques},
  year      = {2017},
  issn      = {1096-4290},
  month     = dec,
  number    = {3},
  pages     = {e21224},
  volume    = {28},
  doi       = {10.1002/mmce.21224},
  publisher = {Wiley},
}

@article{sugimoto_microwave_2025,
	title = {Microwave ablation vs. single-needle radiofrequency ablation for the treatment of {HCC} up to 4 cm: {A} randomized-controlled trial},
	volume = {7},
	issn = {25895559},
	shorttitle = {Microwave ablation vs. single-needle radiofrequency ablation for the treatment of {HCC} up to 4 cm},
	url = {https://linkinghub.elsevier.com/retrieve/pii/S2589555924002738},
	doi = {10.1016/j.jhepr.2024.101269},
	language = {en},
	number = {1},
	journal = {JHEP Reports},
	author = {Sugimoto, Katsutoshi and others},
	month = jan,
	year = {2025},
	pages = {101269},
}

@article{simon_microwave_2005,
	title = {Microwave {Ablation}: {Principles} and {Applications}},
	volume = {25},
	issn = {0271-5333, 1527-1323},
	shorttitle = {Microwave {Ablation}},
	url = {http://pubs.rsna.org/doi/10.1148/rg.25si055501},
	doi = {10.1148/rg.25si055501},
	language = {en},
	number = {suppl\_1},
	journal = {RadioGraphics},
	author = {Simon, Caroline J. and Dupuy, Damian E. and Mayo-Smith, William W.},
	month = oct,
	year = {2005},
	pages = {S69--S83},
}

@article{tu_current_2024,
	title = {Current status and new directions for hepatocellular carcinoma diagnosis},
	volume = {8},
	issn = {25425684},
	url = {https://linkinghub.elsevier.com/retrieve/pii/S2542568424000680},
	doi = {10.1016/j.livres.2024.12.001},
	language = {en},
	number = {4},
	journal = {Liver Research},
	author = {Jinqi Tu and Bo Wang and Xiaoming Wang and Kugeng Huo and Wanting Hu and Rongli Zhang and Jinyao Li and Shijie Zhu and Qionglin Liang and Shuxin Han},
	month = dec,
	year = {2024},
	pages = {218--236},
}

@article{poulou_percutaneous_2015,
	title = {Percutaneous microwave ablation \textit{vs} radiofrequency ablation in the treatment of hepatocellular carcinoma},
	volume = {7},
	issn = {1948-5182},
	url = {http://www.wjgnet.com/1948-5182/full/v7/i8/1054.htm},
	doi = {10.4254/wjh.v7.i8.1054},
	language = {en},
	number = {8},
	journal = {World Journal of Hepatology},
	author = {Poulou, Loukia S},
	year = {2015},
	pages = {1054},
}

@article{servin_simulation_2024,
	title = {Simulation of {Image}-{Guided} {Microwave} {Ablation} {Therapy} {Using} a {Digital} {Twin} {Computational} {Model}},
	volume = {5},
	issn = {2644-1276},
	url = {https://ieeexplore.ieee.org/document/10375319/},
	doi = {10.1109/OJEMB.2023.3345733},
	journal = {IEEE Open Journal of Engineering in Medicine and Biology},
	author = {Servin, Frankangel and others},
	year = {2024},
	pages = {107--124},
}

@article{heshmat_using_2024,
	title = {Using {Patient}-{Specific} {3D} {Modeling} and {Simulations} to {Optimize} {Microwave} {Ablation} {Therapy} for {Liver} {Cancer}},
	volume = {16},
	issn = {2072-6694},
	doi = {10.3390/cancers16112095},
	language = {eng},
	number = {11},
	journal = {Cancers},
	author = {Heshmat, Amirreza and others},
	month = may,
	year = {2024},
	pmid = {38893214},
	pmcid = {PMC11171243},
	pages = {2095},
}

@article{singh_thermal_2020,
	title = {Thermal ablation of biological tissues in disease treatment: {A} review of computational models and future directions},
	volume = {39},
	issn = {1536-8386},
	shorttitle = {Thermal ablation of biological tissues in disease treatment},
	doi = {10.1080/15368378.2020.1741383},
	language = {eng},
	number = {2},
	journal = {Electromagnetic Biology and Medicine},
	author = {Singh, Sundeep and Melnik, Roderick},
	month = apr,
	year = {2020},
	pmid = {32233691},
	pages = {49--88},
}

@article{patel_no-touch_2013,
	title = {No-touch {Wedge} {Ablation} {Technique} of {Microwave} {Ablation} for the {Treatment} of {Subcapsular} {Tumors} in the {Liver}},
	volume = {24},
	copyright = {https://www.elsevier.com/tdm/userlicense/1.0/},
	issn = {10510443},
	url = {https://linkinghub.elsevier.com/retrieve/pii/S1051044313008750},
	doi = {10.1016/j.jvir.2013.04.014},
	language = {en},
	number = {8},
	journal = {Journal of Vascular and Interventional Radiology},
	author = {Patel, Premal A. and Ingram, Liam and Wilson, Iain D.C. and Breen, David J.},
	month = aug,
	year = {2013},
	pages = {1257--1262},
}

@article{Winokur2014,
  author = {Ronald S. Winokur and others},
  title = {Characterization of In Vivo Ablation Zones Following Percutaneous Microwave Ablation of the Liver with Two Commercially Available Devices: Are Manufacturer Published Reference Values Useful?},
  journal = {Journal of Vascular and Interventional Radiology},
  volume = {25},
  number = {12},
  year = {2014},
  pages = {1939-1946.e1},
  issn = {1051-0443},
  doi = {10.1016/j.jvir.2014.08.014},
  url = {https://doi.org/10.1016/j.jvir.2014.08.014}
}

@article{seitel_reinsertion,
author = {Seitel, Alexander and others},
title = {Computer-assisted trajectory planning for percutaneous needle insertions},
journal = {Medical Physics},
volume = {38},
number = {6},
pages = {3246-3259},
doi = {https://doi.org/10.1118/1.3590374},
year = {2011}
}

@misc{reinke2023commonlimitationsimageprocessing,
      title={Common Limitations of Image Processing Metrics: A Picture Story}, 
      author={Annika Reinke and others},
      year={2023},
      eprint={2104.05642},
      archivePrefix={arXiv},
      primaryClass={eess.IV},
      doi={10.48550/arXiv.2104.05642}
}

@article{vaidya_tuning_2022,
	title = {Tuning the {Pennes} {Perfusion} {Rate} to {Model} {Large} {Vessel} {Cooling} {Effects} in {Hepatic} {Radiofrequency} {Ablation}},
	volume = {144},
	issn = {0148-0731, 1528-8951},
	url = {https://asmedigitalcollection.asme.org/biomechanical/article/144/8/084506/1136903/Tuning-the-Pennes-Perfusion-Rate-to-Model-Large},
	doi = {10.1115/1.4053909},
	language = {en},
	number = {8},
	journal = {Journal of Biomechanical Engineering},
	author = {Vaidya, Nikhil and Baragona, Marco and Lavezzo, Valentina and Maessen, Ralph and Veroy, Karen},
	month = aug,
	year = {2022},
	pages = {084506},
}

@phdthesis{audigier:tel-01256010,
  TITLE = {{Computational modeling of radiofrequency ablation for the planning and guidance of abdominal tumor treatment}},
  AUTHOR = {Audigier, Chlo{\'e}},
  URL = {https://theses.hal.science/tel-01256010},
  NUMBER = {2015NICE4071},
  SCHOOL = {{Universit{\'e} Nice Sophia Antipolis}},
  YEAR = {2015},
  MONTH = Oct,
  TYPE = {Theses},
  HAL_ID = {tel-01256010},
  HAL_VERSION = {v1},
}

@misc{Baumgartner2025,
  author       = {Baumgartner, Christian and others},
  title        = {{IT'IS Database for Thermal and Electromagnetic Parameters of Biological Tissues, Version 5.0}},
  year         = {2025},
  doi          = {10.13099/VIP21000-05-0}
}

@article{deshazer_physical_2016,
	title = {Physical modeling of microwave ablation zone clinical margin variance},
	volume = {43},
	copyright = {http://onlinelibrary.wiley.com/termsAndConditions\#vor},
	issn = {0094-2405, 2473-4209},
	url = {https://aapm.onlinelibrary.wiley.com/doi/10.1118/1.4942980},
	doi = {10.1118/1.4942980},
	language = {en},
	number = {4},
	journal = {Medical Physics},
	author = {Deshazer, Garron and Merck, Derek and Hagmann, Mark and Dupuy, Damian E. and Prakash, Punit},
	month = apr,
	year = {2016},
	pages = {1764--1776},
}

@article{cavagnaro_numerical_2015,
	title = {Numerical models to evaluate the temperature increase induced by \textit{ex vivo} microwave thermal ablation},
	volume = {60},
	issn = {0031-9155, 1361-6560},
	url = {https://iopscience.iop.org/article/10.1088/0031-9155/60/8/3287},
	doi = {10.1088/0031-9155/60/8/3287},
	language = {en},
	number = {8},
	journal = {Physics in Medicine and Biology},
	author = {Cavagnaro, M and Pinto, R and Lopresto, V},
	month = apr,
	year = {2015},
	pages = {3287--3311},
}

@article{ge_multi-slot_2018,
	title = {A multi-slot coaxial microwave antenna for liver tumor ablation},
	volume = {63},
	issn = {1361-6560},
	url = {https://iopscience.iop.org/article/10.1088/1361-6560/aad9c5},
	doi = {10.1088/1361-6560/aad9c5},
	language = {en},
	number = {17},
	journal = {Physics in Medicine \& Biology},
	author = {Ge, Mengke and others},
	month = sep,
	year = {2018},
}

@article{boskovic_finite_2023,
	title = {Finite {Element} {Analysis} of {Microwave} {Tumor} {Ablation} {Based} on {Open}-{Source} {Software} {Components}},
	volume = {11},
	copyright = {https://creativecommons.org/licenses/by/4.0/},
	issn = {2227-7390},
	url = {https://www.mdpi.com/2227-7390/11/12/2654},
	doi = {10.3390/math11122654},
	language = {en},
	number = {12},
	journal = {Mathematics},
	author = {Bošković, Nikola and Radmilović-Radjenović, Marija and Radjenović, Branislav},
	month = jun,
	year = {2023},
	pages = {2654},
}

@article{Zhu2013,
  author    = {Zhu, Qing and Shen, Yuanyuan and Zhang, Aili and Xu, Lisa X.},
  title     = {Numerical study of the influence of water evaporation on {Radiofrequency} ablation},
  journal   = {BioMedical Engineering OnLine},
  year      = {2013},
  volume    = {12},
  number    = {1},
  pages     = {127},
  issn      = {1475-925X},
  doi       = {10.1186/1475-925X-12-127},
  url       = {https://doi.org/10.1186/1475-925X-12-127},
  date      = {2013-12-10}
}

@article{Michalewicz1996EvolutionaryAF,
  title={Evolutionary algorithms for constrained engineering problems},
  author={Zbigniew Michalewicz and Dipankar Dasgupta and Rodolphe Le Riche and Marc Schoenauer},
  journal={Computers \& Industrial Engineering},
  year={1996},
  volume={30},
  pages={851-870},
  url={https://api.semanticscholar.org/CorpusID:6368461},
  doi={10.1016/0360-8352(96)00037-X}
}

\end{document}